\documentclass[%
twocolumn,
 amsmath,amssymb,
 aps,
 pre,
]{revtex4}

\usepackage{dcolumn}% Align table columns on decimal poin
\usepackage{lipsum}%
\usepackage{bm}% bold math
\usepackage{graphicx}
\usepackage{comment}
\usepackage{color}
\usepackage{soul}
\definecolor{linkcolor}{rgb}{0,0,0.6} %hyperlink
\usepackage[pdftex,colorlinks=true, pdfstartview=FitV, linkcolor= linkcolor, citecolor= linkcolor, urlcolor= linkcolor, hyperindex=true,hyperfigures=true]{hyperref} %hyperlink% or \documentclass[page-classic]{epl2} for one column style

\newcommand{\average}[1]{\left<{#1}\right>}

\newcommand{\Real}[1]{\Re\left\{{#1}\right\}}
\newcommand{\pq}[1]{\left[{#1}\right]}

\newcommand{\dq}{\dot q}
\newcommand{\om}{\omega}

\newcommand{\dete}{\mathrm{det}}
\newcommand{\tV}{\tilde V}
\newcommand{\teta}{\tilde \eta}

\begin{document}

%\preprint{}

\title { An autonomous out of equilibrium Maxwell's demon for controlling the energy fluxes  produced by thermal fluctuations}

\author{Sergio 
Ciliberto }\email[E-mail me at: ]{sergio.ciliberto@ens-lyon.fr}

\affiliation{ Univ Lyon, Ens de Lyon, Univ Claude Bernard, CNRS, 
Laboratoire de Physique, UMR 5672, F-69342 Lyon, France}

\date{\today}

\begin{abstract}
An autonomous out of equilibrium  Maxwell's demon  is used to reverse the natural direction of the heat  flux between two electric circuits  kept at different temperatures and coupled by the electric thermal noise.  The demon does not process any information, but it achieves its goal by  using  a frequency dependent coupling with the two reservoirs of the system.  There is no energy flux between the demon and the system, but  the total entropy production (system+demon) is positive. The demon can be power supplied by thermocouples. The system and the demon are ruled by equations similar to those of two coupled Brownian particles and of the Brownian gyrator.  Thus our  results pave the way to the application of autonomous out equilibrium Maxwell demons  to coupled nanosystems at different temperatures. 
\end{abstract}

\maketitle

%\section{introduction}
 Nowadays the notion of Maxwell's demon (MD) is generically used to indicate  mechanisms that  allow a system  to execute tasks in apparent violation of the  second law of thermodynamics, such as for example to produce work from a single heat bath and to transfer heat from  cold to  hot sources. To obtain this result  the demon does not exchange energy with the system but it has a positive entropy production rate, which compensate the negative entropy production of the system.  In general the increase in entropy is induced by the fact that the demon needs  to analyze the information that it gathers on the system status\cite{CiliLutz_PT,Parrondo_sagawa}. In experiments this apparent  violation of the second law is obtained  by feedback mechanisms which often  require the  use of external devices such as A/D converters, computers etc.\cite{Toyabe2010,Huard_2017,Roichman_2018,superconductive_2018}.   Several  smart experiments \cite{Raizen_demon_PRL_2008,Pekola_demon_PNAS_2014,Pekola_auton_demon_2015}  have implemented these feedback locally constructing in this way autonomous Maxwell demons, which do not need the use of external devices as the measure and the feedback are performed in the same place. Several autonomous Maxwell demons have been theoretically developed\cite{Aut_dem_Seifert_2013,Shirashi_2015,Boyd_2016,Rosinberg_2016,Demon_Jarzinsky_2019}, but they can be  of difficult practical implementation in several devices such as colloidal particles and mesoscopic electric circuits at room temperature. However it has been recently introduced in ref.\cite{autonmous_demon_PRL} a new paradigm of MD based on an out of equilibrium device, which does not elaborate any information about the system status.  It has been shown that the parameters of this device can be suitably tuned in such a way that it does not exchange energy  (heat or work)  with the system but it has a positive entropy production rate. Thus  it has the two main  requirements of  an autonomous MD and it can be more easily experimentally realized because, in contrast to the commonly used definition of MD, it works without acquiring and analyzing any information about the system status. 
 \begin{figure}[h!]
 	\includegraphics[width=0.8\linewidth]{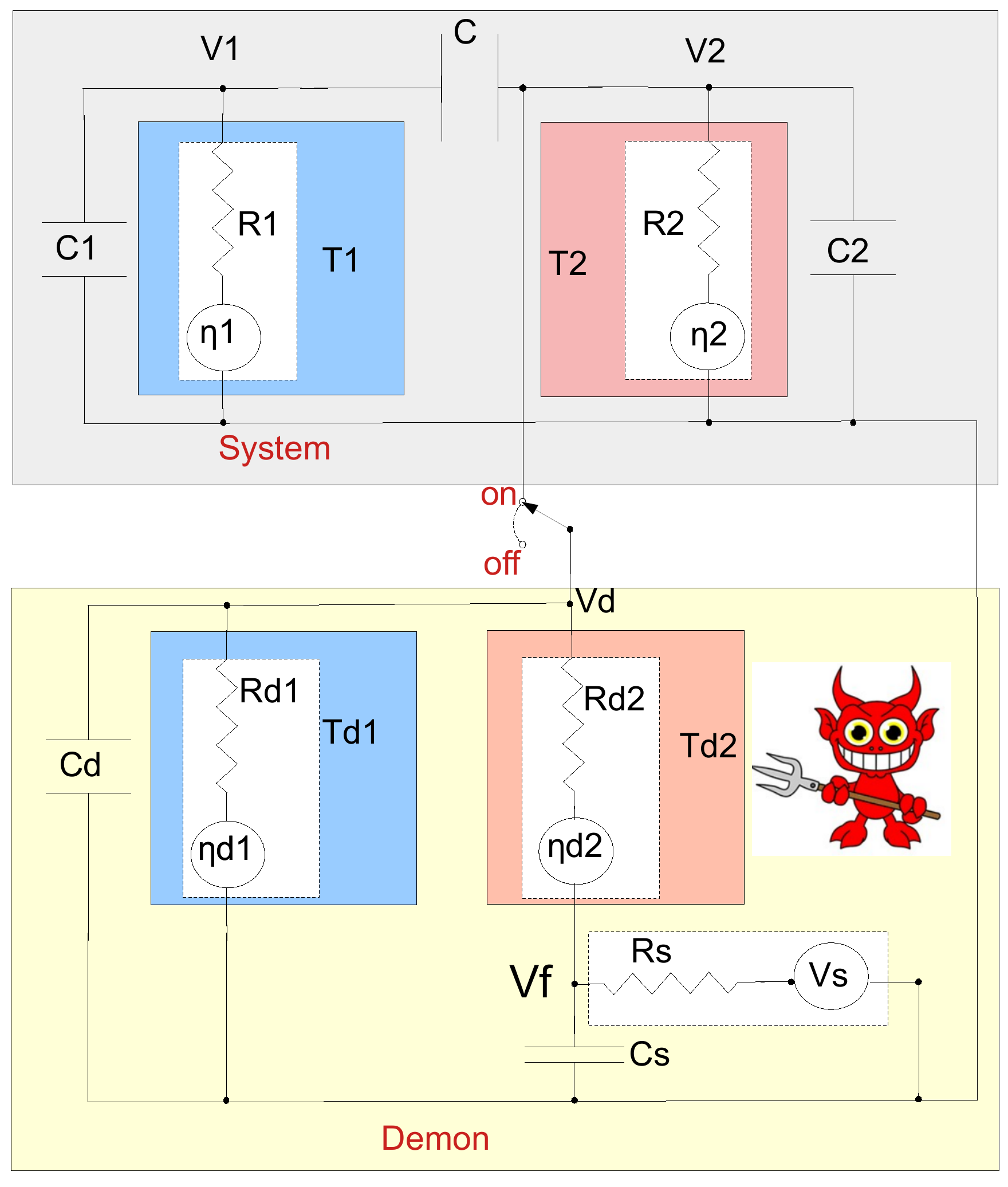}
 	\caption{  Diagram of the system (grey box) and of the demon (yellow box). The system is constituted by the two resistances $R_1$ and $R_2$  kept respectively at temperature $T_1$ and $T_2$, with $T_2\ge T_1$.   They are coupled via the capacitor $C$. The
 		capacitors  $C_1$ and $C_2$ schematize the capacitances of the cables
 		and of the amplifier inputs. The demon (yellow box), is
 		composed by 
 		two resistances ($R_{d1}$ and $R_{d2}$) kept at two different temperatures $T_{d1}$ and $T_{d2}$.Furthermore the  resistance $R_{d2}$ is driven by a voltage generator $V_s$ whose out-put is filtered by the low pas filter composed by the resistance $Rs$ and the capacitance $C_s$.
 		 The four  voltage generators $\eta_{k}$ (k=1,2,d1,d2) represent the Nyquist noise voltages  of the  resistances at the temperatures of the heat baths.
 	}
 	\label{fig:circuit_demon}
 \end{figure}

 We discuss here how to implement an out of equilibrium MD (OEMD) \cite{autonmous_demon_PRL} in electric circuits, which are versatile dynamical systems ruled by coupled Langevin equations \cite{Freitas_circuits_2020,Cohen_electrical}. 
 %We also propose a  simplified version of the original OEMD of ref.\cite{autonmous_demon_PRL}. 
 Thus our study is quite general because it opens the way to the application of OEMDs to coupled nanosystems modeled by Langevin equations. As an example  we will show in this article how  an OEMD can be used 
 to reverse the natural  direction of the heat flux between two electric circuits kept at different  temperatures and coupled by the electric thermal noise \cite{Ciliberto_heat_flux_PRL,Ciliberto_heat_flux_JSTAT}. In fig.\ref{fig:circuit_demon} we sketch the system (gray box)  and the demon (yellow box).
   We chose for the  system this specific circuit because the statistical properties of the heat flux have  been characterized both theoretically and experimentally \cite{Ciliberto_heat_flux_PRL,Ciliberto_heat_flux_JSTAT}. Furthermore it is ruled by the same equations of the Brownian gyrator \cite{Gyrator_2007,Gyrator_2018} and of two Brownian particles coupled by a  harmonic potential and kept at different temperatures\cite{Ciliberto_heat_flux_PRL}, making the result rather  general

%\section{The system}

{\it The system} (grey box in fig.\ref{fig:circuit_demon})   is constituted by two resistances $R_1$ and $R_2$, which are kept at two different temperatures $T_1$ and $T_2\ge T_1$.
% These temperatures are controlled  by thermal baths and $T_2$ is kept fixed at $296K$ whereas $T_1$  can be set at a value  between $296K$ and $88K$ using liquid nitrogen vapor as a circulating coolant.  
In the figure, the two resistances have been drawn with their associated thermal noise generators $\eta_1$ and $\eta_2$, whose power spectral densities  are given by the Nyquist formula $|\tilde \eta_m|^2= 4 k_B R_mT_m$, with $m=1,2$. 
%(see  eqs.\ref{lan1},\ref{lan2} and ref.\cite{Supplementary}).
The coupling capacitance $C$ controls the electrical power exchanged between the resistances  and as a consequence the energy exchanged between the two baths. {No other coupling exists between the two resistances.}. 
The two capacitors  $C_1$ and $C_2$ represents the sum of the circuit and cable capacitances. All the relevant energy exchanges in the system can be derived  by the simultaneous measurements of the voltage $V_m$ ($m=1,2$) across the resistance $R_m$ and the currents $i_m$ flowing through them.
%{All the relevant quantities considered in this paper can be derived by the measurements of $V_1$ and $V_2$, as discussed below}. 
%In the following we will take  $C=100pF,  C_1=680pF, C_t=420pF$ and $R_1=R_2=10M\Omega$, if not differently stated. 
When $T_1=T_2$ the system is in equilibrium and exhibits no net energy flux  between the two reservoirs. 
%This is indeed the condition imposed  by Nyquist to prove his formula, and we use it  {to check all the values of the circuit parameters.} 
The circuit equations can be written in terms of charges $q_m$  flowed through the resistances $R_m$, so  the measured instantaneous currents are  $i_m=\dot q_m $. We make the choice of working with charges because the analogy with a Brownian particle is straightforward 
as $q_m$ is equivalent to  the displacement of the particle $m$ \cite{Cohen_electrical,Ciliberto_heat_flux_PRL,Ciliberto_heat_flux_JSTAT}. 
A  circuit analysis shows that the equations for the charges are: 
\begin{eqnarray}
R_1 \ \dot q_1&=&V_1 - \eta_1, \ \ \rm{and}  \ \ \ R_2\  \dot q_2=  \eta_2-V_2, \label{lan1}
\end{eqnarray}
with
\begin{eqnarray}
V_1&=& \frac{-q_1(C+C_2)+q_2 \ C}{X}  \label{eq_q1}\\
V_2&=& \frac{-q_1 \ C +q_2 \ (C+C_1)} {X}  \label{eq_q2}.
\end{eqnarray}
where  $X=C_2\, C_1+C\, (C_1 \, +C_2) $ and   $\eta_m$ is the Nyquist white noise: $\average{\eta_i(t)\eta_j(t')}=2 \delta_{ij} {k_BT_i R_j} \delta(t-t')$. 
 In ref.\cite{Ciliberto_heat_flux_JSTAT} we have shown that eqs.\ref{lan1}  fully characterize all the thermodynamics properties of the system. 
%
%We use the equation for the dynamics of $V_1$ and $V_2$ from ref.\cite{Ciliberto_heat_flux_JSTAT}. 
%\begin{eqnarray}
%(C_1+C) \dot V_1&=& C \dot V_2 + \frac{1}{R_1}(\eta_1-V_1)\label{lan01},\\
%(C_2+C) \dot V_2&=& C \dot V_1 +\frac{1}{R_2}(\eta_2-V_2)\label{lan02}.
%\end{eqnarray}
%where   $\eta_m$ is the usual white noise: $\average{\eta_i(t)\eta_j(t')}=2 \delta_{ij} {k_BT_i R_j} \delta(t-t')$, 

In this system the work and the heat are defined as 
\begin{eqnarray}
\dot W_m&=&{C \over X}  q_{m'} \dot q_{m} \label{wm}\\ 
\dot Q_m&=&V_m i_m={V_m}\frac{V_m-\eta_m}{R_m}.\label{qm}
\end{eqnarray} 
The quantities  $\dot W_{m}$ is  identified as the thermodynamic work performed by the circuit $m'$  on $m \ne m'$ and $Q_m$ the heat dissipated by the resistance $m$  \cite{Cohen_electrical,Freitas_circuits_2020,Ciliberto_heat_flux_PRL,Ciliberto_heat_flux_JSTAT,Garnier}.
As all the  variables  are fluctuating, the derived quantities $\dot Q_{m}$ and $\dot W_{m}$ fluctuate too. 
In ref.\cite{Ciliberto_heat_flux_PRL} we computed and measured 
the mean heat flux between the two heat baths, which is given by :
\begin{eqnarray}
\average{\dot Q_1}= \ - \average{\dot Q_2} \ = \frac{C^2 k_B (T_2-T_1)}{XY}.
\label{eq:dtQ1} 
\end{eqnarray} 
where $\average{.}$ stands for mean value and  we have introduced the quantity $Y=\pq{(C_1+C) R_1 +(C_2+C) R_2 }$. We use the convention that the heat extracted from a system reservoir is  negative and the heat dissipated is positive.
%The circuits equations can be written in terms of charges.
%Let $q_m$ ($m=1,2$) be the charges that have flowed through the resistances $R_m$, so  the instantaneous current flowing through them is $i_m=\dot q_m $. 
%A  circuit analysis shows that the equations for the charges are: 
%\begin{eqnarray}
%R_1 \dot q_1&=&- q_1\, {C_2 \over X}+  (q_2-q_1){C \over X} + \eta_1  \label{lan1}\\
%R_2 \dot q_2&=&- q_2\, {C_1  \over X}+  (q_1-q_2){C \over X} + \eta_2 \label{lan2}
%\end{eqnarray}
%where we used the transformation 
%\begin{eqnarray}
%q_1&=& (V_1-V_2) \, C + V_1\, C_1  \label{eq_q1}\\
%q_2&=& (V_1-V_2) \, C - V_2\, C_2  \label{eq_q2}.
%\end{eqnarray}
%to pass from eq.\ref{lan1} and \ref{lan2} to the equations eqs.\ref{lan01} and \ref{lan02}

{\it The out of equilibrium demon}
 is sketched in the yellow box of fig.\ref{fig:circuit_demon} and it is
composed by 
two resistances ($R_{d1}$ and $R_{d2}$) kept at two different temperatures $T_{d1}$ and $T_{d2}$ (see Appendix \ref{app:demon}). The two voltage voltage generator $\eta_{d1}$ and $\eta_{d2}$ represent  the Nyquist noise voltages associated to the two resistances at the heat bath temperatures. Furthermore the  resistance $R_{d2}$ is driven by a voltage generator $V_s$ whose out-put is  fltered by the the low pas filter composed by the resistance $Rs$ and the capacitance $C_s$(see Appendix \ref{app:demon_noise}). We notice that demon scheme is similar to that of the system, with a coupling capacitance $C\rightarrow \infty$, on which the driving $V_s$ has been added.  
 To design it,  we followed the  main prescriptions of ref.\cite{autonmous_demon_PRL}: 1) It is out of equilibrium; 2) Either $T_{d1}$ or $T_{d2}$ has to be smaller than $T_1$;  3) it  produces  colored noise, obtained in our case by the source $V_s$ filtered by  $R_s$ and $C_s$ ;4) It is coupled with the two parts of the system on different frequency ranges, specifically high frequencies with subsystem 1 and DC coupled with subsystem 2. 

The choice of $V_s$ is very important in order to simplify the experimental configuration. Indeed $V_s$ can  be  either the thermal fluctuations of $Rs$  with a suitable cut-off imposed by the $R_s C_s$ or  an external driving. Many choices  are possible and  the simplest one is  to use  $V_s=V_f$=constant and $Rs=0$.  In such a way $V_f$ is coupled with $R_2$ only and the thermal noises $\eta_{d1}$ and $\eta_{d2}$ are directly coupled with $R_2$ and high pas filtered for $R_1$ (see Appendix \ref{app:demon_noise})
%Thus the energy exchanges between the demon and each subsystem are different. 
The demon is always out of equilibrium, because, when it is disconnected from the system,  the power  supplied by $V_f$ is entirely  dissipated in the demon resistances producing a mean  heat flux towards the demon heat baths,  even in the case $T_{d1}=T_{d2}$.  
This is a simplified version of the original OEMD of 
ref.\cite{autonmous_demon_PRL} because it requires the use of only one cold source at $T_d$ and a DC signal that can be easily generated by thermocouples making the demon fully autonomous. 
 We will demonstrate 
%We will show 
that this demon can reverse the heat flux of the system in a wide range of parameters with a zero energy flux (heat and work) with the system.
{\it The connection of the  demon to the system}
 changes the current distributions and the energy exchanges. The circuit analysis shows (see Appendix \ref{app:system_curr}), that the currents 
 $\dq_k$ ($k=1,2,d1,d2$) flowing in the resistances $R_k$ are now ruled by the following equations :
\begin{eqnarray}
R_1  \ \dq_{1}&=&V_1-\eta_1 \label{eq:q1}\\
R_2  \ \dq_{2}&=&\eta_2-V_2  \label{eq:q2} \\
R_{d1} \  \dq_{d1}&=&\eta_{d1}-V_2  \label{eq:qd1} \\
R_{d2}   \ \dq_{d2}&=&\eta_{d2}+V_f-V_2  \label{eq:qd2}
\end{eqnarray}    
\begin{eqnarray}
\ V_1&=&  {-(C_t+C)  \  q_1 +   \ C \ q_t\over X_t}   \label{eq:V1C}\\
V_2&=&  {(C_1+C) \  q_t  -  \ C \  q_1 \over X_t}  \label{eq:V2C}
\end{eqnarray}
where  $q_t=(q_2+q_{d2}+q_{d1})$,$C_t=C_2+C_d$.
 and $X_t=C_1C_t+C(C_1+C_t)$.

In order to reduce the number of parameters we consider the case $T_d=T_{d1}=T_{d2}$ and $R_{d1}=R_{d2}$. 
The heat fluxes in the four reservoirs can be  computed  using $\dot Q_k=v_k \dq_k$, where $v_k$ is the potential difference on the resistance $R_k$ (see Appendix \ref{app_system_heat}). 
Introducing the following parameters 
$R_d =R_{d1} R_{d2}/(R_{d1}+R_{d2})$, $R_t=R_{d} R_{2}/(R_{d}+R_{2})$, \\ $Y_t=R_1 \ (C+C_1)+R_t \ (C+C_t)$, $A = {C^2 k_B }/({X_tY_t})$,  \\ $ \average{V_2}= V_t = {V_f  \ R_t / R_{d2} }$,\\ $B=A \ R_t \ (X_t  R_t +R_1 (C_1+C)^2)/ ( R_{d}\ R_{2}\ C^2)$,\\
%\begin{eqnarray}
%R_d&=&({1\over R_{d1}}+{1\over R_{d2}})^{-1}, R_t=({1\over R_{d}}+{1\over R_{2}})^{-1} \\
%Y_t&=&R_1 \ (C+C_1)+R_t \ (C+C_t), \\ A& = &\frac{C^2 k_B }{X_tY_t} \\    \average{V_2}&=& V_t = {V_f  \ R_t \over  R_{d2} } , \\
%B&=&k_B\frac{ X  R_t +R_1 (C_1+C)^2}{X_t \ Y_t  }{ R_t \over R_{d}\ R_{2}}
%\end{eqnarray}
we obtain: 
\begin{eqnarray}
\average{\dot Q_1}&=& A \left( \frac{R_t }{R_2}(T_2-T_1)+  \frac{ R_t}{R_{d}}(T_{d}-T_1) \right) \label{eq:Q1}\\
\average{\dot Q_2} &=& - A  \frac{R_t }{R_2}(T_2-T_1) -  B (T_2-T_{d})\ +{V_t^2 \over R_2} \label{eq:Q2}\\
\average{\dot Q_{d
} } &=&  -  A  \frac{ R_t }{R_{d}}(T_{d}-T_1) -B (T_{d}-T_{2}) +\notag \\ & &+ \ { V_f^2 R_{t} \over  R_{d2}}({1 \over R_{d1}}+ {1 \over R_2}) - {V_t^2 \over R_2} \\
\average{\dot W_f }&=& { V_f^2 \ R_t \over  R_{d2}} ({1 \over R_{d1}}+ {1 \over R_2}) \label{eq:Wd}
\end{eqnarray}
where $\average{\dot Q_{d
} }=\average{\dot Q_{d1
} }+\average{\dot Q_{d2
} }$  is the total heat flux in the demon reservoirs and $\average{\dot W_{f}}$ is  the total  power supplied     by the external generator $V_f$. The total energy balance demon+system is : 
\begin{equation}
\average{\dot Q_{d} }-\average{\dot W_f }+\average{\dot Q_{1} }+\average{\dot Q_{2} }=0
\label{eq:tot_ene}
\end{equation}
These equations allow us to define the conditions for which the demon can  reverse the flow without any energy exchange with the system.

In absence of the demon the heat flux is given by eqs.\ref{eq:dtQ1}, i.e. $\average{\dot Q_2}=-\average{\dot Q_1}<0$. 
Using the demon we want to reverse  this flow  making  $\average{\dot Q_2} >0 $ but  keeping  $\average{\dot Q_1} = -\average{\dot Q_2}$ because an observer, who measures the heat-flux of the system, has to establish that heat flows from the cold to the hot reservoir.   
The  condition  $\average{\dot Q_1} = -\average{\dot Q_2}$ has two  important consequences.   Firstly  it reduces eq.\ref{eq:tot_ene}   to 
 \begin{eqnarray}
 \average{\dot Q_{d} }-\average{\dot W_f} &=& 0 \label{eq:demon_energy_ballance},
 \end{eqnarray}
 which indicates that all the power supplied by $V_f$ is dissipated in the demon reservoirs and not in the system reservoirs.  
 Secondly applying it to eqs.\ref{eq:Q1},\ref{eq:Q2} we find that:
 \begin{eqnarray}
 {V_t^2 \over R_2} &=&   A { R_t \over R_{d} }(T_1-T_{d}) + B \ 
 (T_2-T_{d}) \label{eq:condVt}
 \end{eqnarray}
 Finally using  eq.\ref{eq:condVt} and the  condition   $\average{\dot Q_2} > 0 $ in eq.\ref{eq:Q2}, we  compute the range of $T_d$,  where the  the spontaneous process is reversed, finding:
\begin{eqnarray}
T_{d} \le T_1- \frac{R_d }{R_2}(T_2-T_1)   \label{eq:condTd}
\end{eqnarray}
 
%In the specific case in  which we reverse the initial flux i.e. $\average{\dot Q_2} = {A X_t Y_t \over X \ Y}  (T_2-T_1)$ one finds that 
%\begin{eqnarray}
%T_{d} &=& T_1 - (\frac{X_t Y_t }{X Y}+\frac{R_t }{R_2})(T_2-T_1)  \frac{R_{d}}{ R_t}. \label{eq:condTd}
%\end{eqnarray}
The eqs.\ref{eq:condVt}   and  \ref{eq:condTd} fix the conditions that allows the demon to reverse the system heat flux without heat exchange (eq.\ref{eq:demon_energy_ballance}) between the demon and the system. Eq.\ref{eq:condVt} indicates that the fraction of the  power injected by the demon and  dissipated in $R_2$ ($V_t^2/R_2$ in eq.\ref{eq:Q2}) is compensated  by the heat extracted from the system baths.
We can also prove that thanks to eq.\ref{eq:condVt} the demon does not perform any work on the system. 
Indeed the total work  performed by the demon on the system is :
$ \average{\dot W_{d,s}} =  \average{\dot W_{d,1}} +  \average{\dot W_{d,2}}$
where $\average{\dot W_{d,1}}$ and $\average{\dot W_{d,2}}$ are the works performed on subsystems 1 and 2 respectively. 
These can be computed using equations  eq.\ref{eq:q1} and eq.\ref{eq:q2} in which we see that a ''force'' proportional to $q_{d1}+q_{d2}$ is applied on the two subsystems.  Thus the work per unit time  of these  forces are \cite{Ciliberto_heat_flux_JSTAT}. 
\begin{eqnarray}
\average{\dot W_{d,1}}&=& \frac{C}{X_t}\average{\dq_{1} \ (q_{d1}+q_{d2})} \\
\average{\dot W_{d,2}}&=& -\frac{C_1+C} {X_t}\average{\dq_{2} \ (q_{d1}+q_{d2})}
\end{eqnarray}
From these two works (computed in Appendix \ref{sec:compute_W_d_s}) 
 we obtain for the total work :
%\begin{eqnarray}
 %\average{\dot W_{d,1}} &=&  {A \ R_t^2 \over  R_d} \left( {  (T_2-T_1) \over  \ R_2} - { (T_1-T_d) \over  \ R_d} %\right) \\
%\average{\dot W_{d,2}} &=& -B \ (T_2-T_d)-{A \ R_t^2 \over   R_d } \left({T_2-T_1 \over R_2}+ {T_1-T_d \over %R_2}\right)+  {V_t^2 \over R_2}
%\end{eqnarray}
\begin{eqnarray}
\average{\dot W_{d,s}}=  - A { R_t \over  R_d } (T_1-T_d) -B \ (T_2-T_d)+  {V_t^2 \over R_2}
\label{eq:total_work}
\end{eqnarray}
We clearly see that if the condition on  $V_t$ (eq.\ref{eq:condVt}) is verified then $ \average{\dot W_{d,s}}=0$, i.e. no work is done by the demon on the system. Thus eq.\ref{eq:condVt} and eq.\ref{eq:demon_energy_ballance} insure  that total energy flux from the demon to the system is  zero.
% and the circuit in the yellow box in fig.\ref{fig:circuit_demon}  behaves as a demon. %for the range of $T_d$ fixed by eq.\ref{eq:condTd}. 
%%Indeed eq.\ref{eq:condVt} indicates that the fraction of the  power injected by the demon and  dissipated in $R_2$ ($V_t^2/R_2$ in eq.\ref{eq:Q2}) is compensated  by the heat extracted from the system baths and the net energy flux from the  demon to the system is zero.   
\begin{figure}[t!]
		\includegraphics[width=0.9\linewidth]{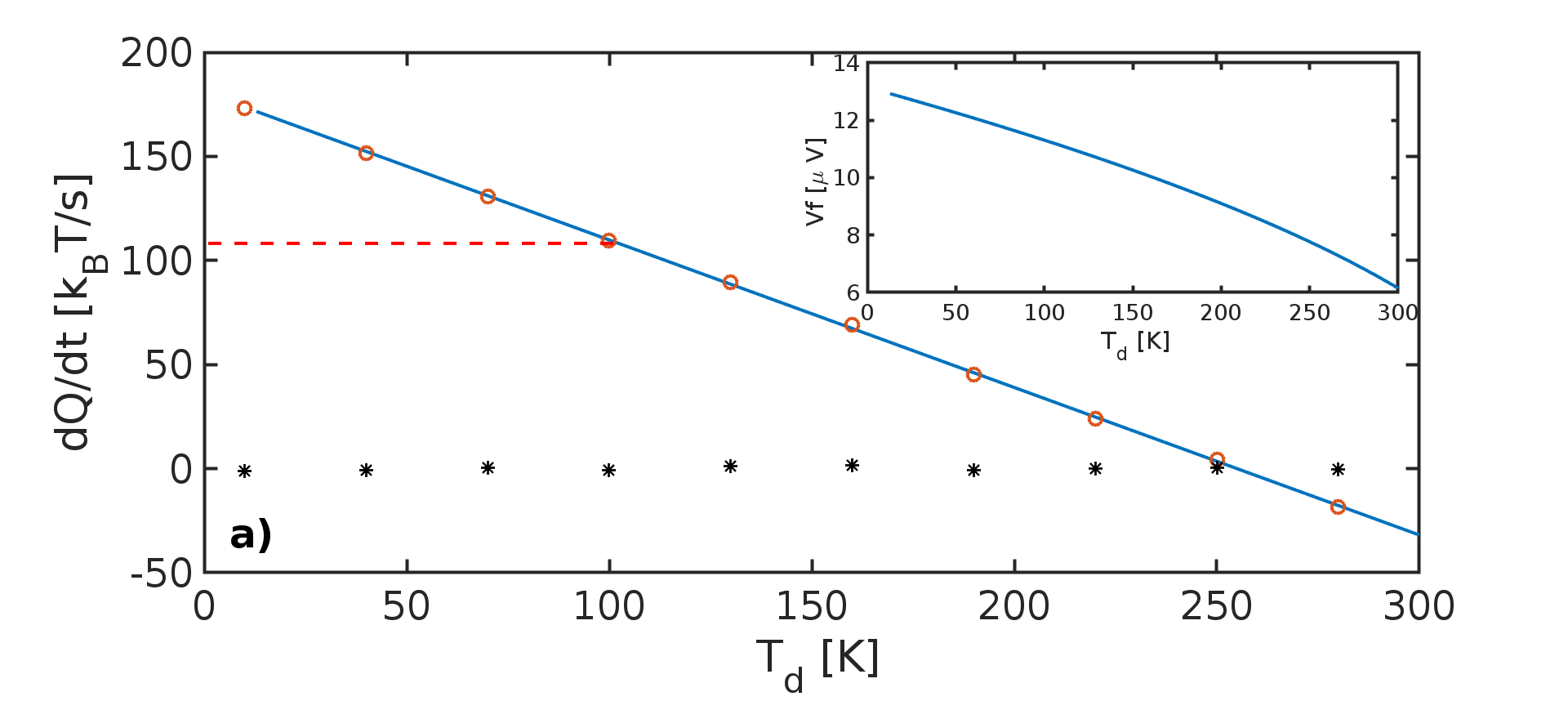}\\ \includegraphics[width=0.9\linewidth]{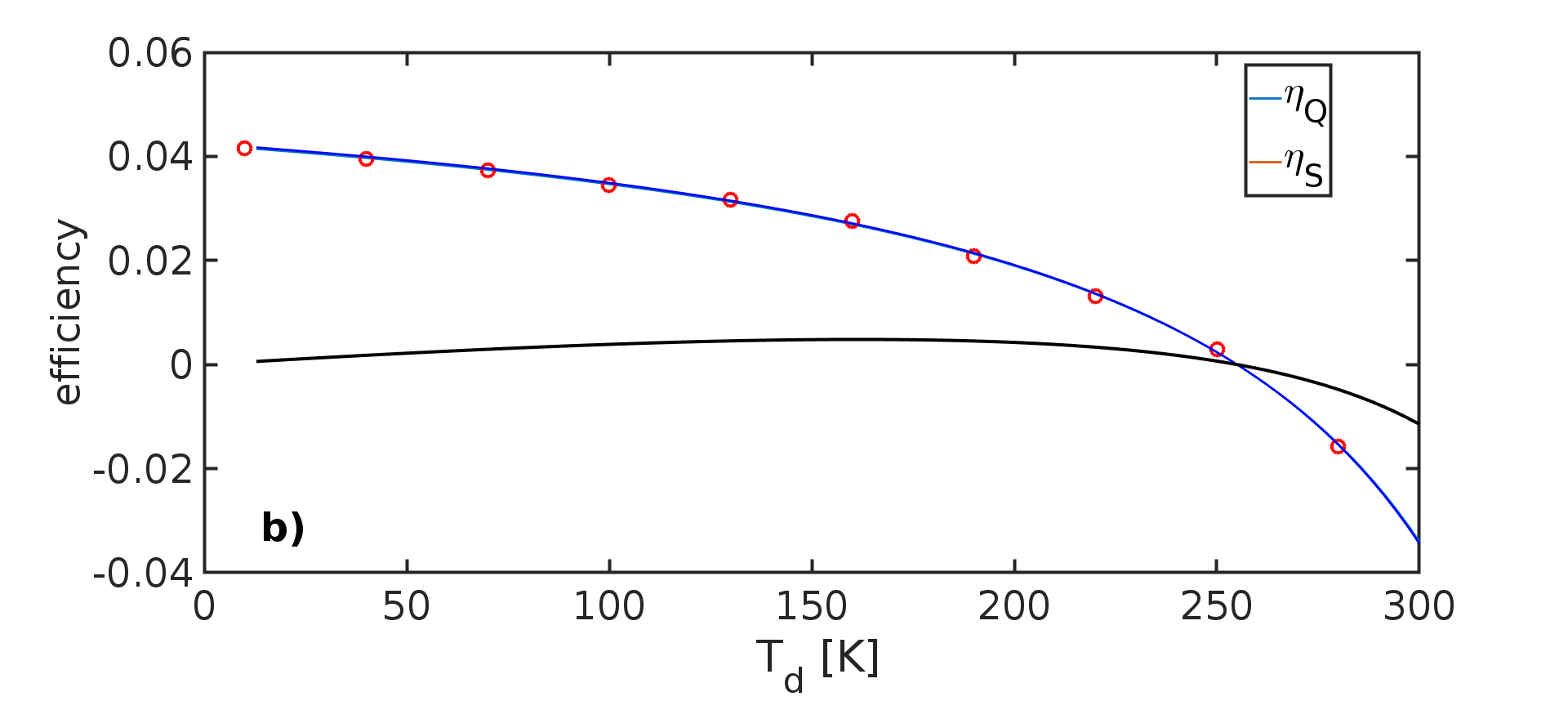}
		\caption{a) Heat fluxes as a function of the demon temperature $T_d$ :$\average{\dot Q_2}$ (blue line)  computed when the demon is on,  using eqs.\ref{eq:Q1},..,\ref{eq:Wd} and the condition for $V_t^2$ eq.\ref{eq:condVt}; $\average{\dot Q_1}$ (horizontal red dashed line) computed (eq.\ref{eq:dtQ1}) when the demon is ''off'' ;  $\average{\dot Q_2}$ (red circles) and  $\average{\dot Q_2}+\average{\dot Q_1}$ (black stars)  obtained from the direct numerical simulation of eqs.\ref{eq:q1},..,\ref{eq:qd2}. b) Demon efficiencies as a function of $T_d$: $\eta_s$ (black line) and $\eta_Q$ (blue line, red circles)  computed from  eqs.\ref{eq:Q2},\ref{eq:Wd},\ref{eq:etas} (continuos lines) and  obtained from the direct numerical simulations of eqs.\ref{eq:q1},..,\ref{eq:qd2} (red circles). 
			The parameters  used to compute the curves in a) and b) are:  $T_1=300$K, $T_2=450$K,$C=1$nF, $C_1=C_2=100$pF, $R_1=R_2=10\rm{M}\Omega$;$R_d=3\rm{M}\Omega$;$Cd=50$pF.}
\label{fig:simul}
\end{figure}

However the demon produces entropy and the total entropy production rate  $\average{\dot S}$ is positive in spite of the fact that the system entropy production rate \\ $\average{\dot S_s}=\average{\dot Q_2}\left( {1 }/{T_{2}}- {1 }/{T_{1}} \right)$ is negative, because $\average{\dot Q_2}>0$ and $T_2>T_1$ when the demon is "on". The total entropy production rate is
\begin{eqnarray}
\average{\dot S}% &=&  
%\frac{\average{\dot Q_d} }{T_{d}}+ \frac{\average{\dot Q_1} }{T_{1}}+ \frac{\average{\dot Q_2} }{T_{2}} =   \\
 &=&  \frac{\average{\dot Q_d} }{T_{d}}+ \average{\dot Q_2}\left( \frac{1 }{T_{2}}- \frac{1 }{T_{1}} \right)  \label{eq:entropy_rate}
\end{eqnarray}
To show that $\average{\dot S}>0$, we start  by taking into account that
% that $R_t<R_d<R_{d2}$ and $R_t<R_2$ we deduce that  $\average{\dot W_d} > {V_t^2 \over R_2}$   
 $\average{\dot Q_d} = \average{\dot W_f}$  (see eq.\ref{eq:demon_energy_ballance}) and that $\average{\dot W_f}> {V_t^2 \over R_2}$ because as we said ${V_t^2 \over R_2}$ is a fraction of the total power injected into the system by the demon source (computed Appendix \ref{section_other_relations}). 
  Furthermore as we want $\average{\dot Q_{2}}>0$ then from eq.\ref{eq:Q2} we have that $V_t^2/R_2>\average{\dot Q_2}$ as the other terms are negative because $T_2>T_1>T_d$. 
As a consequence \\
$ \average{\dot Q_d} /T_{d} > \average{\dot Q_d} (1/T_1-1/T_2) >  \average{\dot Q_2} (1/T_1-1/T_2) $
and we find  $\average{\dot S}>0$. 

These results on the effect of the demon on the system can be checked by comparing the heat fluxes computed  from eqs.\ref{eq:Q1},\ref{eq:Q2} with those  obtained  by the direct numerical integration of eqs.\ref{eq:q1}-\ref{eq:qd2}, where the four  $<\dot Q_k>=<v_k\dq_k>$ are  directly computed using Stratonovich integrals. This comparison is done  
using for the system components (i.e.$R_1,R_2,C_1,C_2,C$) the values of  the experiment of ref.\cite{Ciliberto_heat_flux_PRL,Ciliberto_heat_flux_JSTAT}.  For the demon, we chose for $C_d$ a typical wiring value  and we fixed $R_d<R_2$  for having a reasonable range of $T_d$ (see eq.\ref{eq:condTd}).  All the components and temperatures values  are indicated in the caption of fig.\ref{fig:simul}. In fig.\ref{fig:simul}a) the horizontal red dashed line indicates $|<\dot Q_2>|$ at $T_2-T_1=150$K computed  from eq.\ref{eq:dtQ1} when the demon is ''off''. When the demon is ''on'' the value of $\dot Q_2$ computed from eqs.\ref{eq:Q2} and that obtained from direct numerical simulation agree. Most importantly for $T_d<250$K the heat flows from the cold to the hot thermal  bath. The values of $V_t$ necessary for implementing the demon conditions (eq.\ref{eq:condVt})
are plotted in the inset indeed for these values of $V_t$ we see that $<\dot Q_2>+<\dot Q_1>=0$ in the numerical simulation. It is important to notice that the necessary $V_t$ is of the order of a few microvolts meaning that it can be easily obtained by two thermocouples coupled with a cold and an hot bath for example $T_1$ and $T_2$. 

{\it The Demon efficiency} can be defined in two ways. As the demon does not exchange any work and heat with the system then the efficiency can be defined in terms of  entropy production rates, which has been used in other  contexts \cite{Verley_2014,Polettini_2015,Derivaux_2019}. Another way to define efficiency is in terms of the energy fluxes. Specifically these efficiencies  are :
\begin{equation}
\eta_s=-\ {\dot S_s\over \dot S_d} \ \ \rm{and} \ \ \eta_Q= {<\dot Q_2> \over <\dot W_f>}. \label{eq:etas}
\end{equation}
We see that $\eta_s$ is  the ratio between the entropy of the non spontaneous process divided by the entropy of the spontaneous process whereas $\eta_Q$ is the ratio between the reversed heat flux in the system divided by the work performed by the demon to achieve the goal. These two quantities are plotted in fig.\ref{fig:simul}b) as function of $T_d$, and we observe that $\eta_s <1\%$ and $\eta_Q < 4\%$, i.e. in order to achieve its goal the demon has to do a lot of work with a very large entropy production rate.  

To conclude,  we have simplified the original idea of autonomous OEMD because we use a demon with a single bath and a DC forcing (powered by thermocouples) instead of two baths with colored noise as in ref.\cite{autonmous_demon_PRL}. We have demonstrated that this autonomous OEMD can be  applied to electric circuits in order to reverse the spontaneous heat processes with no energy exchange  between the system and demon. The latter 
% i.e. the  demon does not perform any work on the system and there is no  total  energy flux between the system and the demon. 
 has  a small efficiency and a  very large positive entropy production rate that largely compensates the negative entropy production rate of the system. Our results are very general because they are based on four coupled Langevin equations, which model not only electric circuits, but a lot of micro and nano-systems.  Thus this article paves the way to the general applications of OEMDs to the control of these mesoscopic systems. We have chosen this configuration as a proof of principle  but other complex circuits can of course be implemented.% because it is the simplest but  one can do the same using for example resonant circuits.
% which are governed by second order Langevin equations making the results even more general. 

\acknowledgements 
We acknowledge useful discussion with R.S.Whitney. This work has been supported by the FQXi foundation on grant number FQXi-IAF19-05 ''Information as a fuel in colloids and superconducting quantum circuits ''

\bibliographystyle{apsrev4-1}
\bibliography{maxwell_demon.bib}
%\nocite{*}
\newpage
\begin{widetext}
\appendix
\section{Several details about  the   demon } \label{app:demon}
In this section we describe the out of equilibrium demon, composed by 
two resistances ($Rd_1$ and $Rd_2$) kept at two different temperatures $Td_1$ and $Td_2$. The two voltage generators $\eta d_1$ and $\eta d_2$ corresponding to the Nyquist noise of the two resistances at the temperatures of the heat baths. Furhermore the  resistance Rd2 is driven by a voltage generator $V_s$ whose out-put is  fltered by the the low pas filter composed by the resistance $Rs$ and the capacitance $C_s$.
\subsection{The demon as a colored noise generator} \label{app:demon_noise}
 The dynamics of the demon can be obtained by writing the Kirchhoff laws for the points $V_f$ and $V_d$ :
\begin{eqnarray}
{V_s-V_f \over R_s}- \dot V_f   \  C_f-{V_f+ \eta_{d2} -V_d \over R_{d2}} & = & 0     \\
{V_f+\eta_{d2}-V_d \over R_{d2}}- \dot V_d \ C_d-{V_d - \eta_{d1} \over R_{d1}} -i_d & = & 0
\end{eqnarray}
where  $i_d$ is the current flowing between the system and the demon when the latter  is ''on''. 
From these equations we get: 
\begin{eqnarray}
\tau_f \ \dot V_f=-V_f + {R_f \over R_s}  V_s+ {R_f \over R_{d2}} ( V_d - \eta_{d2}) \\
\tau_d \ \dot V_d =-V_d + \left( {R_d \over R_{d2}}  V_f -i_d\ R_d\right)+ \xi_d \\
\end{eqnarray}
where we use :
\begin{eqnarray}
R_d & = &{R_{d1} \  R_{d2} \over R_{d1}+R_{d2} },   \   \   \tau_d=R_d C_d,   \  \  
R_f = {R_s \ R_{d2} \over R_{d2}+R_s },  \  \  \   \  \tau_f=R_fC_s \ \  \ \rm{and} \  \  \   \  \xi_d=\eta_{d1} {R_d \over R_{d1}} +\eta_{d2} {R_d \over R_{d2}}. 
\end{eqnarray}
As $V_s$ is not necessarily the thermal noise of $R_s$, it can be a very large external driving that allows the demon to work, we can simplify this circuit if we assume 
that $R_{d2}>>R_s$
\begin{eqnarray}
\tau_f \ \dot V_f &=&-V_f +   V_s  \\
\tau_d \ \dot V_d &=&-V_d - \left( {R_{d} \over R_{d2}}  V_f -i_d \ R_{d}\right)+ \xi_d
\label{eq:V_demon}
\end{eqnarray}
These equations  show that the demon produces colored noise for the system as it is needed to control the system heat flux. 
\subsection{The heat flux in the demon switched off} \label{app:demon_flux}
When $V_s=0$ and $i_d=0$ the heat transfers between the  two reservoirs of the demon can be obtained   from eqs.\ref{eq:dtQ1} for $C\rightarrow \infty$. Thus the heat transfers in the demon are: 
\begin{eqnarray}
\average{\dot Q_{d1}}&=& \frac{k_B (T_{d2}-T_{d1})}{(R_{d1}+R_{d2})C_{d}}.
\label{dtQd1} \\
\average{\dot Q_{d2}}&=& \frac{k_B (T_{d1}-T_{d2})}{(R_{d1}+R_{d2})C_{d}}. \label{dtQd2}
\end{eqnarray} 
When there is an external driving one has also to consider the work performed by  the external source $V_f$ still in the case $i_d=0$,i.e. with the demon is not  connected to the system. 
We can use for $Vs$ either thermal fluctuations of $Rs$  with a suitable cut-off imposed by the $R_s C_s$ or  an external driving. In the simplest version we make the choice  to use a $V_f=V_s$ constant and $Rs= 0$. In such a case the heat dissipated by the two resistances is: 
\begin{eqnarray}
\average{\dot Q_{d1}}&=& \frac{k_B (T_{d2}-T_{d1})}{(R_{d1}+R_{d2})C_{d}}+ {V_f^2 R_{d1}\over (R_{d1}+R_{d2})^2 }  
\label{dtQd1_W} \\
\average{\dot Q_{d2}}&=& \frac{k_B (T_{d1}-T_{d2})}{(R_{d1}+R_{d2})C_{d}} + {V_f^2 R_{d2}\over (R_{d1}+R_{d2})^2 } \label{dtQd2_W}
\end{eqnarray}

and doing the energy balance we have
\begin{equation}
\average{\dot Q_{d1}}+\average{\dot Q_{d2}}=\ {V^2 \over (R_{d1}+R_{d2}) }= \ \average{\dot  W_f}
\end{equation}
where $\average{\dot  W_f}$ is the power injected into the demon by the voltage generator.
where we use the convention  that the heat  dissipated in the bath and the work performed on the system are positive. 
Thus the demon is always out of equilibrium even at $T_{d1}=T_{d2}$

\section{The  currents in the system+demon circuit} \label{app:system_curr}

We now connect the demon to the system and this connection changes the current distributions. We write the equations of the circuits in terms of charges. The relationships between currents and charges are 
\begin{eqnarray}
0&=&-\dq_1-\dq_{c1}+\dq_c,  \ \ \ \dq_2-\dq_{ct}-\dq c+\dq_{d2}+\dq_{d1}=0, \notag \\   
% \dq_c &=& ({\dq_{ct}\over C_t}-{\dq_{c1}\over C_1})C  \  
%\rm{with}  \ \       V_1  C_1 = q_{c1} \ \rm{and} \ V_2  C_t = qc_{t} 
\dq_c &=& (V_2-V_1)C,  \  V_1  C_1 = q_{c1} \ \rm{and} \ \ V_2  C_t = q_{ct} 
\end{eqnarray}
where  $\dq_k$ ($k=1,2,d1,d2$) are the currents flowing in the resistances $R_k$. Futhermore $\dot q_c, \dot q_{c1}$ and  $\dot q_{ct} $ are  the currents flowing respectively in the capacitors  $C,C_1$ and $C_t=C_2+C_d$.
% which is the parallel of $C_2$ with $C_d$.  
Solving the system for $q_{c1}$ and $q_{ct}$ we find :
\begin{eqnarray}
V_1={q_{c1} \over C_1}&=&  {-(C_t+C)  \  q_1 +   \ C \ q_t\over X_t}   \label{eq:V1C}\\
V_2={q_{ct} \over C_t}&=&  {(C_1+C) \  q_t  -  \ C \  q_1 \over X_t}  \label{eq:V2C}
\end{eqnarray}
where  $q_t=(q_2+q_{d2}+q_{d1})$ and $X_t=C_1C_t+C(C_1+C_t)$.
we can now solve for the four currents flowing in the 2 system resistances and the two demon resistances specifically 
\begin{eqnarray}
R_1 \ \dq_{1}&=&V_1-\eta_1 = {-(C_t+C) \  \  q_1 +   \  C \ (q_2+q_{d2}+q_{d1})\over X_t} -\eta_1 \label{eq:Bq1}\\
R_2 \  \dq_{2}&=&\eta_2-V_2 = {-(C_1+C) \  \  (q_2+q_{d2}+q_{d1}) +   \  C \ q_1 \over X_t} +\eta_2 \label{eq:Bq2} \\
R_{d1}  \ \dq_{d1}&=&\eta_{d1}-V_2 = {-(C_1+C) \  \  (q_2+q_{d2}+q_{d1}) +   \  C \ q_1 \over X_t} +\eta_{d1} \label{eq:Bqd1} \\
R_{d2}  \  \dq_{d2}&=&\eta_{d2}+V_f-V_2 = {-(C_1+C) \  \  (q_2+q_{d2}+q_{d1}) +   \  C \ q_1 \over X_t} +\eta_{d2}+V_f \label{eq:Bqd2}
\end{eqnarray} 
We write the equations for the charges because the connection with Brownian particles is straightforward ($q$ correspond to the dispacement). Furthermore it is  more clear to understand the amount of work performed by the demon on the system and  the amount of dissipated heats  in the various reservoirs. 
As the resistances of the demon are just in parallel to $R_2$ the  equations \ref{eq:Bq1},.,\ref{eq:Bqd2} can be reduced to:
\begin{eqnarray}
R_1 \ \dq_{1}&=& -{(C_t+C) \over X_t}\  \  q_1 +   \  {C \over X_t} \ q_t -\eta_1 \label{eq:q1t}\\
R_t \  \dq_{t}&=&  -{(C_1+C) \over X_t} \  \  q_t +   \  {C  \over X_t} \  q_1 + \ \eta_t+ \ V_t \label{eq:q2t} 
\end{eqnarray}
where we define 
\begin{eqnarray}
q_t  & = &(q_2+q_{d2}+q_{d1}), \ \ \ \ R_t=(1/R_2+1/R_{d2}+1/R_{d1})^{-1},\ \  \rm{and}  \notag,\\  \ \ V_t &=& V_f \ R_t/ R_{d2}, \ \ \ \eta_t=(\eta_2/R_2+\eta_{d1}/R_{d1}+\eta_{d2}/R_{d2})R_t.
\end{eqnarray}
Using eqs.\ref{wm},\ref{qm} we can compute the heat and the work of the different parts of the circuit. 
We can also use eqs. \ref{eq:dtQ1} to compute the total heat exchanged between the reservoir at $T_1$ with the reservoirs of resistance $R_2,R_{d1},R_{d2}$. Defining 
$T_t=R_t( T_2 /R_2+T_{d1} /R_{d1}+T_{d2} /R_{d2})$ and $Y_t=R_1 \ (C+C_1)+R_t \ (C+C_t)$, we get : 
\begin{eqnarray}
\average{\dot Q_1}&=& \frac{C^2 k_B (T_t-T_1)}{X_tY_t}.
\label{eq:dtQ1t} \\
\average{\dot Q_t}&=& \frac{C^2 k_B (T_1-T_t)}{X_tY_t} \label{eq:dtQ2t}
\end{eqnarray} 
where the contribution of the work performed by $V_t$ cancels out in the stationary regime (see appendix \ref{section_other_relations}). Eq. \ref{eq:dtQ1t}  can be decomposed in the various contributions to the heat fluxes from the three reservoir at $(T_2,T_{d1}$ and $T_{d2})$ :
\begin{eqnarray}
\average{\dot Q_1}&=& \frac{C^2 k_B }{X_t Y_t} \left( \frac{R_t }{R_2}(T_2-T_1)+  \frac{R_t}{R_{d1}}(T_{d1}-T_1)+ \frac{R_t}{R_{d2}}(T_{d2}-T_1) \right)
\label{eq:dtQ1d}
\end{eqnarray}

\section{Calculation of the heat fluxes} \label{app_system_heat}

By the definition eq.\ref{qm} the heat fluxes in the resistances $k=1,2,d1,d2$ are given by:  
\begin{eqnarray}
\average{\dot Q_k }&=& \average{{ v_k} {( v_k-\eta_k)\over R_k }}.\label{eq:Qm}
\end{eqnarray} 
where ${v_k}= V_k- V_f  \ \delta_{k,d2}$ (with $V_{d1}=V_{d2}=V_2$) is the potential difference on the resistance $R_k$. 
The mean values can be evaluated by using the Fourier transforms $\tilde {v_k}$ of $v_k$, which can be written as : 
 \begin{eqnarray}
\average{\dot Q_k }&=&\int_0^\infty \Real{ {\frac{\tilde v_k^*}{R_m}(\tilde v_k-\tilde \eta_k)} } \ \frac{d\omega}{2 \ \pi}.\label{eq:Qmfourier}
\end{eqnarray}   
To evaluate this integral we take into account that the spectral density of $\tilde \eta_k$ is $|\tilde \eta_k|^2=4 k_B T_k R_k$ and that $\Real{ \eta_k \ \eta_{k'}^*}=0$ because different  noise sources are uncorrelated
We have already computed $\average{\dot Q_1}$ in eq.\ref{eq:dtQ1t} and eq.\ref{eq:dtQ1d} using the eqs.\ref{eq:dtQ1}. We need to estimate  all the heat fluxes in the demon reservoirs  and in the reservoir 2 by computing  $V_1$ and $V_2$ given by eqs.\ref{eq:V1C} and \ref{eq:V2C}. Their values depend on $q_1,q_t$ which can be computed solving the system of eqs.\ref{eq:q1t} and \ref{eq:q2t} in Fourier space. 
 \begin{eqnarray}
 Z_1 \ \tilde q_1 - a \ \tilde q_t & =& -\tilde \eta_1 \\
 -a \ \tilde q_1 + Z_t \ \tilde q_t &=& \tilde \eta_t + V_t \ \sqrt{2\pi} \delta(\om)
 \end{eqnarray}   
where $Z_1=(C_t+C)/X_t+i \omega R_1  $, $Z_t=(C_1+C)/X_t+i \omega \ R_t $ and $a=C/X_t$. We find 
 \begin{eqnarray}
\tilde q_1&=&{-\teta_1 \ Z_t + \eta_t a   \over \mathrm{det} } \\
\tilde q_t&=&{\teta_t \ Z_1 - \eta_1 a   \over \mathrm{det} } \\
\dete &=& Z_1 \ Z_t-a^2={ (1-\om^2 R_1 R_t X_t) + i\om \ Y_t \over X_t } 
\end{eqnarray}   
Inserting these Fourier transform in  eqs.\ref{eq:V1C} and \ref{eq:V2C}, we get:
 \begin{eqnarray}
\tV_1 & =& \frac{\tilde \eta_1 [1+i \om \ R_t (C_t+C)] +\ \tilde \eta_t \ i\om R_1 C}{X_t\  \dete} \label{eq:ftV1}  \\
& & \notag \\
\tV_2 & =& \frac{\tilde \eta_t [1+i \om \ R_1 (C_1+C)] +\ \tilde \eta_1 \ i\om R_t C}{X_t\  \dete}+V_t \ \sqrt{2\pi} \delta(\om) \label{eq:ftV2} 
\end{eqnarray}  
 We see that 
  \begin{eqnarray}
  |\tV_1|^2&=&\frac{|\teta_1|^2  \pq{1+(\om \ R_t (C_t+C))^2} +  |\tilde \eta_t|^2(\om R_1 C)^2}{ (1-\om^2 R_1 R_t X_t)^2 + (\om \ Y_t)^2 } \\
  |\tV_2|^2&=&\frac{|\teta_t|^2  \pq{1+(\om \ R_1 (C_1+C))^2} +  |\tilde \eta_1|^2(\om R_t C)^2}{ (1-\om^2 R_1 R_t X_t)^2 + (\om \ Y_t)^2 } + V_t^2 {2\pi} \delta(\om)
\end{eqnarray}
\begin{eqnarray}
\average{V_1^2}&=&\int_{0}^{\infty} |\tV_1|^2  \frac{d\omega}{2 \ \pi}= \frac{|\teta_1|^2 (X_t R_1 R_t +R_t^2 (C_t+C)^2+ |\teta_t|^2 R_1^2 C^2 }{ 4 Y_t X_t \ R_1 \ R_t} \\
\average{V_2^2}&=&\int_{0}^{\infty} |\tV_2|^2  \frac{d\omega}{2 \ \pi}= \frac{|\teta_t|^2 (X_t R_1 R_t +R_1^2 (C_1+C)^2)+ |\teta_1|^2 R_t^2 C^2 }{4 \ Y_t X_t \ R_1 \ R_t}+V_t^2
\end{eqnarray}
where we used the integrals:
\begin{equation}
\int_{0}^{\infty} \frac{Y_t}{{ (1-\om^2 R_1 R_t X_t)^2 + (\om \ Y_t)^2 }}\  \frac{d\omega}{2 \ \pi}=\frac {1}{4} 
\ \ \ \rm{and} \ \ \ 
\int_{0}^{\infty} \frac{Y_t \ \omega^2}{{ (1-\om^2 R_1 R_t X_t)^2 + (\om \ Y_t)^2 }} \ \frac{d\omega}{2 \ \pi}=\frac {1}{4\ R_1 R_t X_t }
\end{equation}

% \begin{eqnarray}
% \tilde {\dot Q}_t&=&   \Real{ \tV_2^* \frac{(\tV_2-\teta_t- V_t \sqrt{2\pi}\delta(\om))}{R_t}}=\frac{|\tV_2|^2}{R_t}- \Real{\frac{|\tilde \eta_t|^2 [1-i \om \ R_1 (C_1+C)]\dete}{{R_t} \ X_t |\dete|^2}}= \notag  \\  
%&=&  \frac{|\tV_2|^2}{R_t} -  \frac{|\tilde \eta_t|^2 \pq{(1-\om^2 R_1 R_t X_t)+ \om^2 Y_t \ R_1 \ (C_1+C)}}{{R_t} \ \pq{(1-\om^2 R_1 R_t X_t)^2 + (\om \ Y_t)^2} } \\
%\average{\dot Q_t}&=&  \frac{|\teta_t|^2 (X_t R_1 R_t +R_1^2 (C_1+C)^2)+ |\teta_1|^2 R_t^2 C^2 - |\teta_t|^2 Y_t \ R_1 \ (C_1+C) }{4 \ Y_t X_t \ R_1 \ R_t^2} = \notag \\
%&=& -\frac{k_B C^2 (T_t-T_1)}{X_t \ Y_t }
%\end{eqnarray} 
%where we used the definition of $Y_t$ and that $|\teta_m|^2=4k_B \ T_m R_m$ where $T_t=(T_{d1}/R_{d1}+ T_{d2}/R_{d2}+ T_{2}/R_{2}) R_t$.
%\begin{eqnarray}
%\tilde {\dot Q}_1&=&   \Real{ \tV_1^* \frac{(\tV_1-\teta_1)}{R_1}}=\frac{|\tV_1|^2}{R_1}- \Real{\frac{|\tilde \eta_1|^2 [1-i \om \ R_t  (C_t+C)]\dete}{{R_1} \ X_t |\dete|^2}}= \notag  \\  
%&=&  \frac{|\tV_1|^2}{R_1} -  \frac{|\tilde \eta_1|^2 \pq{(1-\om^2 R_1 R_t X_t)+ \om^2 Y_t \ R_t \ (C_t+C)}}{{R_1} \ \pq{(1-\om^2 R_1 R_t X_t)^2 + (\om \ Y_t)^2} } \\
%\average{\dot Q_1}&=&  \frac{|\teta_1|^2 (X_t R_1 R_t +R_t^2 (C_t+C)^2)+ |\teta_t|^2 R_1^2 C^2 - |\teta_1|^2 Y_t \ R_t \ (C_t+C) }{4 \ Y_t X_t \ R_1^2 \ R_t} = \notag \\
%&=& \frac{k_B C^2 (T_t-T_1)}{X_t \ Y_t }=-\average{\dot Q_t}
%\end{eqnarray} 
We can now compute the other heat fluxes : 
\begin{eqnarray}
\tilde {\dot Q}_2&=&   \Real{ \tV_2^* \frac{(\tV_2-\teta_2)}{R_2}}=\frac{|\tV_2|^2}{R_2}- \Real{\frac{|\tilde \eta_2|^2 R_t [1-i \om \ R_1 (C_1+C)]\dete}{{R_2}^2 \ X_t |\dete|^2}}= \notag  \\  
&=&  \frac{|\tV_2|^2}{R_2} -  \frac{|\tilde \eta_2|^2 \ R_t \pq{(1-\om^2 R_1 R_t X_t)+ \om^2 Y_t \ R_1 \ (C_1+C)}}{{R_2}^2 \ \pq{(1-\om^2 R_1 R_t X_t)^2 + (\om \ Y_t)^2} } \\
& & \notag \\
& & \notag \\
\average{\dot Q_2}&=& \int_0^\infty \tilde {\dot Q}_2 \ \frac{d\omega}{2 \ \pi} = \notag \\ &=& \frac{|\teta_t|^2 \ R_2 \ (X_t R_1 R_t +R_1^2 (C_1+C)^2)+R_2 |\teta_1|^2 R_t^2 C^2 - |\teta_2|^2 R_t \ Y_t \ R_1 \ (C_1+C) }{4 \ R_2^2 \  Y_t  X_t \ R_1 \ R_t} = \notag \\
&=&  \frac{(|\teta_t|^2 \ R_2 -|\teta_2|^2 R_t)\ R_1(X_t  R_t +R_1 (C_1+C)^2)+  R_t^2 C^2 (|\teta_1|^2 R_2- |\teta_2|^2 R_1) }{4 \ R_2^2 \  Y_t  X_t \ R_1 \ R_t} + {V_t^2 \over R_2} \notag \\
&=& -\frac{k_B C^2 }{X_t \ Y_t }(T_2-T_1) \frac{R_t}{R_2}- k_B\frac{ X_t  R_t +R_1 (C_1+C)^2}{X_t \ Y_t  R_2} \left( {R_t \over R_{d1}}(T_2-T_{d1})+{R_t \over R_{d2}}(T_2-T_{d2}) \right)+{V_t^2 \over R_2}
\end{eqnarray} 

\begin{eqnarray}
\tilde {\dot Q}_{d2}&=&   \Real{ (\tV_2^*-V_f \sqrt{2\pi}\delta(\om)) \frac{(\tV_2-\teta_{d2}-V_f \sqrt{2\pi}\delta(\om))}{R_{d2}}} \notag \\ &=&\frac{|\tV_2|^2}{R_{d2}}- \Real{\frac{|\tilde \eta_{d2}|^2 R_t [1-i \om \ R_1 (C_1+C)]\dete}{{R_{d2}}^2 \ X_t |\dete|^2}}+ {V_t(V_t-V_f) \over R_{d2}}2 \pi \delta(\om)= \notag  \\  
&=&  \frac{|\tV_2|^2}{R_{d2}} -  \frac{|\tilde \eta_{d2}|^2 \ R_t \pq{(1-\om^2 R_1 R_t X_t)+ \om^2 Y_t \ R_1 \ (C_1+C)}}{{R_{d2}}^2 \ \pq{(1-\om^2 R_1 R_t X_t)^2 + (\om \ Y_t)^2} }+ {(V_t-V_f)^2 \over R_{d2}} 2\pi\delta(\om)\\
& & \notag \\
& & \notag \\
\average{\dot Q_{d2}}&=& \int_0^\infty \tilde {\dot Q}_{d2} \ \frac{d\omega}{2 \ \pi} = \notag \\ &=&  \frac{|\teta_t|^2 \ R_{d2} \ (X_t R_1 R_t +R_1^2 (C_1+C)^2)+R_{d2} |\teta_1|^2 R_t^2 C^2 - |\teta_{d2}|^2 R_t \ Y_t \ R_1 \ (C_1+C) }{4 \ R_{d2}^2 \  Y_t  X_t \ R_1 \ R_t}+{(V_t-V_f)^2 \over R_{d2}} = \notag \\
&=&  \frac{(|\teta_t|^2 \ R_{d2} -|\teta_{d2}|^2 R_t)\ R_1(X_t  R_t +R_1 (C_1+C)^2)+  R_t^2 C^2 (|\teta_1|^2 R_2- |\teta_{d2}|^2 R_1) }{4 \ R_{d2}^2 \  Y_t  X_t \ R_1 \ R_t} + {(V_t-V_f)^2 \over R_{d2}}=\notag \\
&=& -A  \ (T_{d2}-T_1) \frac{R_t}{R_{d2}}- B \frac{ R_d}{ R_{d2}} \left( {R_2 \over R_{d1}}(T_{d2}-T_{d1})+(T_{d2}-T_2) \right)+{(V_t-V_f)^2 \over R_{d2}} \\
\rm{where }& & \notag \\
A&=& \frac{k_B C^2 }{X_t \ Y_t } \ \ \  \rm{and} \ \ \ B=k_B  \frac{ R_t(X_t  R_t +R_1 (C_1+C)^2)}{X_t \ Y_t R_2 R_{d}} \label{eq:def_A_and_B}
\end{eqnarray}
Finally using the same method we get: 
\begin{eqnarray}
\average{\dot Q_{d1}}&=&  -A  \ (T_{d1}-T_1) \frac{R_t}{R_{d1}}- B \frac{ R_d}{ R_{d1}} \left( {R_2 \over R_{d2}}(T_{d1}-T_{d2})+(T_{d1}-T_2) \right)+{V_t^2 \over R_{d1}} 
\end{eqnarray}
\section{The work performed by the demon on the system} \label{sec:compute_W_d_s}
We follow refs. \cite{Ciliberto_heat_flux_PRL,Ciliberto_heat_flux_JSTAT} to compute the power injected by the demon  on $R_1$ and $R_2$ with $R_{d1}=R_{d2}$ and $T_{d1}=T_{d2}$ 
\begin{eqnarray}
\average{\dot W_{d,1}}&=& \frac{C}{X_t}\average{\dq_{1} \ (q_{d1}+q_{d2})}= \int_{0}^{\infty} { \Real{{\tV_1^*-\teta_1^* \over R_1} \ {\teta_{d} -\tV_2\over R_d \ i \om}}} {d\om  \over 2 \pi} =  \\
&=& -\frac{C}{X_t} \int_{0}^{\infty} \left[|\teta_1|^2 R_t \ C-|\teta_t|^2 R_1 \ C  + (1-\om^2 R_1 \ R_t \ X_t)R_t\ C \ (-|\teta_1|^2+|\teta_{d}|^2 R_1/R_d) \over R_1 R_d |det|^2 X_t^2 \right]{d\om  \over 2 \pi}= \\ &=&- k_B{C^2 R_t \over X_t \ Y_t \ R_d}(T_1-T_t) 
= {A \ R_t^2 \over  \ R_d \ R_2}(T_2-T_1)-{A \   R_t^2 \over \ R_d^2}(T_1-T_d) \label{eq_Wd1}
\end{eqnarray}
where we used  eqs\ref{eq:ftV1}, \ref{eq:ftV2}, $T_t=R_t(T_2/R_2+T_d/R_d)$ and eqs.\ref{eq:def_A_and_B}.
\begin{eqnarray}
\average{\dot W_{d,2}}&=& - \frac{(C_1+C)}{X_t} \average{\dq_{2} \ (q_{d1}+q_{d2})}= - \frac{(C_1+C)}{X_t}\int_{0}^{\infty} { \Real{{\teta_2^*  -\tV_2^*\over R2} \ {\teta_{d} -\tV_2\over R_d \ i \om}}} {d\om  \over 2 \pi} + {V_t^2 \over R_2}= \\
&=& - {(C_1+C) \over X_t \ R_2 \ R_d} \int_{0}^{\infty} { ({|\teta_2|^2 R_t / R_2 }-{|\teta_d|^2 R_t / R_d}) [Y_t-R1 (C_1+C) (1-\om^2 R_1 \ R_t \ X_t) ]\over X_t^2 |det|^2}{d\om  \over 2 \pi} + {V_t^2 \over R_2} \\ &=&  -k_B {(C_1+C) R_t\over X_t \ R_2 \ R_d} (T_2-T_d)+ {V_t^2 \over R_2}.
\end{eqnarray}
Adding a subtracting $A / R_t^2 T_1/(  R_2 \ R_d)$ from the last expression,taking into account the definition of $A$ and $B$ (eq.\ref{eq:def_A_and_B}),  that 
$(C_1+C)(C_t+C)=X_t+C^2$ and that $ k_B (C_1+C) R_t Y_t/ (X_t \ R_2 \ R_d Y_t)=B+A / R_t^2 /( R_2 \ R_d)$
we get 
\begin{eqnarray}
\average{\dot W_{d,2}} &=& -B \ (T_2-T_d)-{A \ R_t^2 \over  R_2\ R_d } (T_2-T_1)-{A \  R_t^2 \over R_2\ R_d } (T_1-T_d)+  {V_t^2 \over R_2}
\end{eqnarray}
The total work performed by the demon on the system is 
\begin{eqnarray}
\average{\dot W_{d,s}}&=&\average{\dot W_{d,1}}+\average{\dot W_{d,2}}= \\
&=&  - A { R_t \over  R_d } (T_1-T_d) -B \ (T_2-T_d)+  {V_t^2 \over R_2}
 \label{eq:total_work}
\end{eqnarray}
If the condition on the demon potential and temperature  eqs.\ref{eq:condTd} and  \ref{eq:condVt}  are satisfied then the last equation imposes that $\average{\dot W_{d,s}}=0$, i.e. the demon does not perform any work on the system.

\section{The power dissipated by the DC currents only} \label{section_other_relations}
The power dissipated in the system and demon resistances by the DC external generator $V_f$ are:
\begin{eqnarray}
P_{1} &=& 0, \ \ \  
P_{2}=V_t^2/R_{2}, \ \ \ 
P_{d1}=V_t^2/R_{d1}, \ \ \ \ \ \ 
P_{d2}=(V_f-V_t)^2/R_{d2} 
\end{eqnarray}
with $V_t=\average{V_2}$.
Taking into account that $V_t=V_f \ R_t /R_{d2}$
we get the total power supplied  by the generator $V_f$ 
\begin{equation}
\average{\dot W_f}=P_{2}+P_{d1}+P_{d2}= {V_f (V_f-V_2) \over R_{d2} }
\end{equation} 
The dissipation of the demon is $P_d=P_{d1}+P_{d2}$ thus $P_2=\dot W_d -P_d$. This has been used in the text for computing the total entropy

Concerning the work of $V_t$ in eq.\ref{eq:q2t} we clearly see that for the node 2,  $\average{\dq_{t}}=0$ if we consider only the DC current imposed by $V_f$. This means that $V_t \   \average{\dq_{t}}=0$ and there is no contribution to the heat flux $\dot{Q}_t$ which is the heat exchanged by the equivalent $R_t$ circuit with the subsystem 1. 
%Notice that For the stationary case 
%$V_t= {(C_1+C) \over X_t} \  \  q_t$ in eq. \ref{eq:q2t} which is coherent with eq.\ref{eq:V2C} when $q_1=0$ which is %the case in absence of noise. 
\end{widetext}
\end{document}